\documentclass[twocolumn,prl,showpacs]{revtex4}
\usepackage{graphicx}
\usepackage{psfig,epsfig}
\usepackage{dcolumn}
\usepackage{bm}

\begin{document}

\title{Glass glass transition and new dynamical singularity\\ points
in an analytically solvable $p$-spin glass like model}
\author{Antonio Caiazzo, Antonio Coniglio, Mario Nicodemi}
\affiliation{Dipartimento di Scienze Fisiche, Universit\'{a} ``Federico II'',
INFM-Coherentia and INFN Napoli, via Cintia, I-80126 Napoli, Italy
}


\begin{abstract}
We introduce and analytically study a generalized $p$-spin glass like
model that captures some of the main
features of attractive glasses, recently found by Mode Coupling
investigations, such as a
glass/glass transition line and dynamical singularity points
characterized by a logarithmic time dependence of the relaxation. The model
also displays features not predicted by the Mode Coupling scenario
that could further describe the attractive glasses behavior, such
as aging effects with new dynamical singularity points
ruled by logarithmic laws or the presence of a glass spinodal line.
\end{abstract}

\pacs{64.70.Pf, 05.50.+q 64.60.Ht}

\maketitle

Recent Mode Coupling Theory (MCT) \cite{gotze} investigations discovered
\cite{MCT} a new kind of glasses, the so called attractive glasses, 
characterized by the presence of a glass/glass transition line where 
higher order dynamical singularity points are located. As MCT 
calculations triggered intense experimental researches \cite{exp1,exp2}, 
their limit is that, by definition, they can correctly describe only 
the region outside the glass.
In this paper, we introduce a schematic $p$-spin glass like model 
exhibiting a glass/glass transition which we can analytically investigated 
to establish the proper scenario of the glassy phase. This leads to new 
testable predictions on the aging dynamics in that region, where new types 
of singularities are found.

In MCT attractive glasses the potential
between the particles presents attraction for
some range besides the hard core repulsion \cite{MCT}. Its volume
fraction - temperature phase diagram shows two branches of the
glass transition line, one due to the repulsive part of the
potential, extending to high temperatures for volume fractions
near the close packing value, and one related to the attraction,
extending to low volume fractions up to meet the liquid/gas
spinodal. For sufficiently narrow widths of the potential well one
of the two branches interestingly enters in the glass region
giving rise to a glass/glass transition line, ending in a higher
order dynamical singularity point. 
This special point in the MCT framework is characterized by
a logarithmic time dependence of the relaxation.
The theory applies
to attractive colloids; actually, some evidence
of the MCT results has been found by recent experiments
\cite{exp1,exp2}. Nevertheless, for some aspects, the approach
used in \cite{MCT} turns out to be inadequate. It is known in fact
that MCT is valid only for equilibrium dynamics and
cannot be trivially extended inside the glassy region,
where the glass/glass line itself is located and
aging phenomena are present.
Another problem arises when the
phase coexistence interferes with the glass transition line. In
this case MCT does not allow a self-consistent description of the
structural relaxation and the gas/liquid spinodal line is
usually obtained {\it a priori}, not taking into
account the presence of the glassy phase. The purpose of the
present work is to begin to cure these inconsistencies. Exploiting
the strong analogies between glassy systems and the class of
'discontinuous' spin glasses noticed in the last years
\cite{review}, we introduce a schematic solvable model providing a
coherent picture of the glass/glass line, its off-equilibrium dynamics
and singularity points.

The model consists of a combination of two (or more)
$p$-spin models \cite{N1,SZ,CD} diluted through density variables \cite{ANS}.
We analytically solve its Langevin dynamics and for comparison with
MCT inside the glassy region we also use the Mode Coupling
approximation. We show that:
(A) The glass/glass transition line does not coincide with that predicted by
MCT but follows a different curve (Figs. 1,2).
(B) Along the glass/glass line  (see Fig. 1)
in the early regime the stationary autocorrelation function $\Phi(t-t')$
exhibits, like in MCT, a power law relaxation except at its endpoint
$A_3$ where is logarithmic.
(C) In the long time regime, where MCT predicts a plateau, we find
an interesting aging dynamics. In particular,
two more dynamical singularity points, $B_3$
and $B_3'$ (one for each kind of glass, see Fig.1) 
characterizes the relaxation in
the aging regime. On one side of the glass/glass line
from $C$ to $B_3$ (resp. $B_3'$ on the other side, see inset in Fig. 1)
the long times relaxation is a power law, except right at $B_3$
where it is logarithmic (analogously for $B_3'$). From $B_3$ (resp. $B_3'$)
to $A_3$, the aging behavior is spin glass-like \cite{review}.
(D) Finally, at the $A_4$ singularity (i.e., in the case where
$A_3$ coincides with point $C$ in Fig. 1)
we show that the logarithmic decay of the correlation function in the
stationary regime $\Phi \left(t-t^{\prime }\right)$
is followed by an aging regime
ruled again by a logarithmic law.
(E) We calculate in a consistent way the density-temperature phase diagram
and the spinodal lines (see Fig. 1) and find
they are influenced by the presence of the glassy phase.

{\em The Model.}
We consider a lattice gas model, where particles ($n_i=0,1$) interact in
groups of $p$ via their spins ($s_i=\pm 1$) through quenched random
couplings $J_{i_1\cdots i_p}$, with zero mean and variance $p!J_p^2/\left(
2N^{p-1}\right) $, and also in groups of $r$ through (nonrandom) coupling
constants $K_r$ (typically below we take two or three values of $p$ and one
value of $r$).
The mean field Hamiltonian is given by:
\begin{eqnarray}
H &=&-\sum_p\sum_{i_1<\cdots <i_p}J_{i_1\cdots i_p}s_{i_1}n_{i_1}\cdots
s_{i_p}n_{i_p}  \nonumber \\
&&-\sum_r\frac{r!K_r}{2N^{r-1}}\sum_{i_1<\cdots <i_r}n_{i_1}\cdots
n_{i_r}-\mu \sum_in_i  \label{ham}
\end{eqnarray}
where $\mu $ is the chemical potential. The spherical version of
the model (\ref{ham}) is here considered: we introduce two new spin
variables $s_{ai}$ ($a=1,2$%
), defined by $s_i=s_{1i}$, $n_i=\left( s_{1i}s_{2i}+1\right) /2$, and
enforce for these ones spherical constraints ($\sum_is_{ai}^2=N$) \cite{BEG}.
To exclude the possibility of continuous transitions, not related to
a glass transition, the restriction $p>2$ is taken \cite{N1}.
The two spin fields $s_{ai}$ ($a=1,2$) evolve via usual Langevin
equations with thermal noises $\xi _{ai}\left( t\right) $, having zero mean
and variance $\left\langle \xi _{ai}\left( t\right) \xi _{bj}\left(
t^{\prime }\right) \right\rangle =2T\delta _{ab}\delta _{ij}\delta \left(
t-t^{\prime }\right) $.
Due to its quadratic nature the model is exactly solvable. It has a
gas, a liquid and a glassy phases, respectively called $P^{-}$,
$P^{+}$ and $G$ (see Fig. 1).
We focus here on the dynamics, and in particular on the 
correlation function $C\left(
t,t^{\prime }\right) =\sum_{ia}\left\langle s_{1i}\left( t\right)
s_{ai}\left( t^{\prime }\right) \right\rangle /\left( 2N\right)$, 
the related response function $G\left( t,t^{\prime}\right)$ 
and the density 
$d\left( t\right) =C\left( t,t\right) $. Standard functional
techniques \cite{SZ,CL} allow to derive the following
dynamical equations for these quantities after a rapid
quench at $t=0$ from high temperature ($t\ge t^{\prime }$):
\begin{eqnarray}
\frac{\partial C\left( t,t^{\prime }\right) }{\partial t} &=&
-\frac{z\left(
t\right) }2C\left( t,t^{\prime }\right) +2TG\left( t^{\prime },t\right)
\nonumber \\
&&+\frac 12\int_0^{t^{\prime }}du\,\varphi ^{\prime }\left( C\left(
t,u\right) \right) G\left( t^{\prime },u\right)   \nonumber \\
&&+\frac 12\int_0^tdu\,\varphi ^{\prime \prime }\left( C\left( t,u\right)
\right) G\left( t,u\right) C\left( u,t^{\prime }\right)   \label{C-eq}
\end{eqnarray}
\vspace{-0.5cm}
\begin{eqnarray}
\frac{\partial G\left( t,t^{\prime }\right) }{\partial t} &=&-\frac{z\left(
t\right) }2G\left( t,t^{\prime }\right) +\frac{\delta \left( t-t^{\prime
}\right) }2  \nonumber \\
&&+\frac 12\int_{t^{\prime }}^tdu\,\varphi ^{\prime \prime }\left( C\left(
t,u\right) \right) G\left( t,u\right) G\left( u,t^{\prime }\right)
\label{G-eq}
\end{eqnarray}
\vspace{-0.5cm}
\begin{equation}
d^{\prime }\left( t\right) =\left( z\left( t\right) +2\mu +2\chi
^{\prime }\left( d\left( t\right) \right) \right) \left( 1-d\left( t\right)
\right) -T  \label{d-eq}
\end{equation}
Here we define $\varphi \left( x\right) =\left( 1/2\right) \sum_pJ_p^2x^p$%
, $\chi \left( x\right) =\left( 1/2\right) \sum_rK_rx^r$ and
$z\left( t\right) =z_1\left( t\right) -\mu -\chi ^{\prime }\left(
d\left( t\right) \right) $ with $z_1\left( t\right) =z_2\left(
t\right) $ the two time-dependent Lagrange multipliers related to
the spherical constraints (the prime stands for the
first derivative). The above equations completely define the
dynamical model, given the initial density $d\left( 0\right)$. In
the following we consider the large times limit, where one-time
quantities reach stationary values: $d=d\left( t\rightarrow \infty
\right) $ and $z=z\left( t\rightarrow \infty \right) $.

\begin{figure}[h!]
\begin{center}
\includegraphics*[scale=.43]{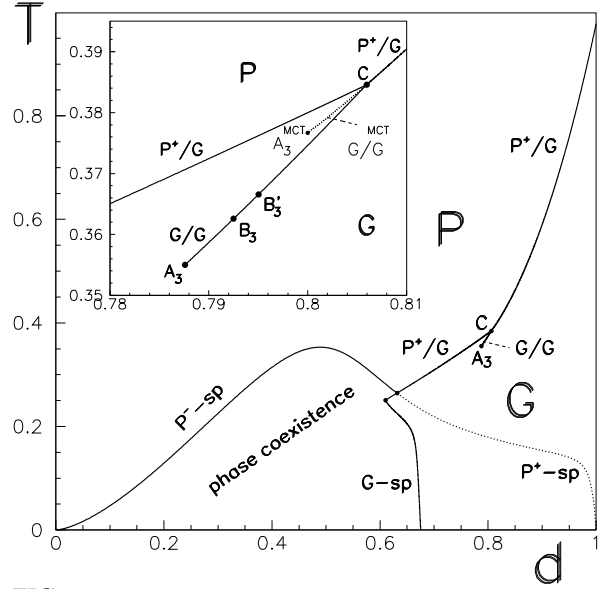}
\end{center}
\vspace{-0.9cm}
\caption{\footnotesize
The model dynamical phase diagram in the density-temperature plane 
(the parameters are here: $p_1=3$, $p_2=11$, 
$J_3^2=.74$, $J_{11}^2=4$ and $r=12$, $K_{12}=-12$).
One can
have a fluid ($P$) phase consisting of gas ($P^{-}$) and liquid ($P^{+}$), or
a glass ($G$) phase. Continuous lines depict the liquid/glass ($%
P^{+}/G$) and glass/glass ($G/G$) transition lines, the spinodal of the
fluid ($P^{-}$-sp and $P^{+}$-sp) and that of the glass ($G$-sp).
The {\bf inset} shows an
enlargement of the region around the $G/G$ transition line. The dotted
line represents the incorrect $G/G$ line predicted by MCT.
On the $G/G$ line (full line from $C$ to $A_3$), 
along with its endpoint $A_3$, there are new
dynamical singularity points, $B_3,B_3^{\prime }$,
characterized by a logarithmic relaxation. Also
the crossing point $C$ between the two branches of
the $P^{+}/G$ transition line is shown.
}
\vspace{-0.5cm}
\end{figure}

{\em MCT calculations.}
Assuming equilibrium dynamics, we have that $C=C(t-t')$ and
$G=G(t-t')$ $\forall t,t'$ and the Fluctuation-Dissipation
theorem (FDT) holds: $TG\left( \tau \right) =-C^{\prime }\left( \tau \right)$
(with $\tau =t-t^{\prime }$).
Under these hypotheses, Eqs. (\ref{C-eq},\ref{G-eq}) yield the long time
schematic MCT equation \cite{nota1} for the correlator
$\Phi \left( \tau \right) =C\left( \tau \right)/d$,
\begin{equation}
\tau _0\frac{d\Phi \left( \tau \right) }{d\tau }+\Phi \left( \tau \right)
+\int_0^\tau d\tau ^{\prime }m\left( \tau -\tau ^{\prime }\right) \frac{%
d\Phi \left( \tau ^{\prime }\right) }{d\tau ^{\prime }}=0  \label{MCT-eq}
\end{equation}
where $m\left( \tau \right) =\varphi ^{\prime }\left( C\left( \tau \right)
\right) d/T^2$. In the standard MCT notation \cite{gotze} the kernel $%
m\left( \tau \right) $ is usually written as $m\left( \tau \right) =F\left(
\Phi \left( \tau \right) ,v\right) $ with $F\left( f,v\right) =\sum_nv_nf^n$%
; this allows to establish a mapping relating $T,d,J_p$ and
the usual MCT control parameters $v_n$, given
by $v_{n}=pJ_p^2d^p/\left( 2T^2\right)$ where $n=p-1$.
If $p_1,p_2,\ldots $ denote the values assumed by $p$,
we deal with the so-called $F_{p_1-1,p_2-1,\ldots }$ MCT model.
The non-ergodicity parameter $f=\Phi \left( \tau
\rightarrow \infty \right) $ is obtained by solving the bifurcation
equation $Z\left( f\right)=1$ (i.e., the $\tau\rightarrow\infty$ limit
of Eq. (\ref{MCT-eq})), where
\begin{equation}
Z\left( f\right) =\frac 1{1-f}-F\left( f,v\right) =\frac 1{1-f}-\sum_p%
\frac{pJ_p^2d^p}{2T^2}f^{p-1}  \label{zf}
\end{equation}
The inequality $1\leq Z\left(\Phi \left( \tau \right) \right) $,
deriving from the decreasing character of $\Phi\left( \tau \right)$,
implies that $f$ is the largest solution of the bifurcation equation
\cite{gotze}.
The conditions $Z\left( f_c\right) =1$ and $Z^{\prime }\left(
f_c\right) =0$ determine the transition line in the plane $d-T$ (see Fig. 1).
The $P^{+}/G$ transition, where $f$ jumps from zero to
a nonzero value, $f_c$, corresponds to singularity points of type
$A_2$ \cite{gotze}.
MCT Eq. (\ref{MCT-eq}) describes the relaxation when this critical line
is approached from the $P^+$ phase:  $\Phi (\tau)$ decays in two steps
characterized by power laws with critical exponents determined
by the parameter $\lambda =1-\left( 1-f_c\right) ^3Z^{\prime
\prime }\left( f_c\right) /2<1$ \cite{gotze}.
Inside the glassy region, for a set of values of the parameters $v_n$,
an other line of $A_2$ singularities appear
(dotted line in the inset of Fig. 1)
where two non zero solutions for the dynamical order parameter, $f$,
are allowed.
This line defines the glass-glass ($G/G$) transition, where
a first glass, $f_{c1}$, transforms discontinuously into
a second one, $f_{c2}$ \cite{gotze,MCT,gotze1}. Its endpoint, where
$\lambda =1$, i.e., $Z^{\prime\prime }\left( f_c\right) =0$,
results to be
an higher order singularity $A_3$ (with $f_{c1}=f_{c2}$).
To have a $G/G$ line in our model, at least two values of $p$ are needed
\cite{CCNfuture}:
as an example, for $p_1=3$ the lowest other possible
value is $p_2=11$ (the case of Fig.s 1,2).

{\em The glassy region and the new G/G line.} Eq.(\ref{MCT-eq}),
i.e., MCT, is correct only out of the glassy phase (for $f=0$), where the
equilibrium dynamics assumption holds. Thus, the above results on
the $G/G$ line are not correctly described by this theory. Within
our model, we discuss now the more complex non-equilibrium
equations governing the glassy phase, the correct position of the
G/G transition and the new aging behavior along the G/G line.
In the glassy phase, as in the usual $p$-spin glass model,
the relaxation of $\Phi(t,t')$ can be split in two
parts \cite{review,CD,CL}: in the first, describing the so called FDT regime 
($t\simeq t^{\prime }$), 
$\Phi(t,t')$ is a function, $\Phi_{FDT}(\tau)$, of the times difference
$\tau=t-t'$;  the second part for $t\gg t^{\prime }$ corresponds to
an aging regime described by a function $\Phi_{AG}(t,t')$.
In the FDT regime, i.e., the approach of the correlator to the
plateau $f_c$, $\Phi_{FDT}(\tau)$ satisfies the following
equation:


\begin{equation}
\tau _0\frac{d\Phi \left( \tau \right) }{d\tau }+Z\Phi \left( \tau \right)
+\left( 1-Z\right)
+\int_0^\tau d\tau ^{\prime }m\left( \tau -\tau ^{\prime }\right) \frac{%
d\Phi \left( \tau ^{\prime }\right) }{d\tau ^{\prime }}=0  \label{MCT-eq1}
\end{equation}
where the existence of a well defined aging solution fixes $Z$ to
the minimum value (not necessarily equal to $1$) satisfying the
stability requirement $Z \left( f\right) \leq Z\left( \Phi_{FDT} \left(
\tau \right) \right) $, i.e., the one corresponding to the
marginal stability condition $Z^{\prime }\left( f\right) =0$
\cite{CD,CL,KH} (see Fig. 2). Note that Eq.(\ref{MCT-eq1}) coincides
with (\ref{MCT-eq}) only for $Z=1$.
The $G/G$ transition line results thus to be modified
with respect to MCT
and located by the conditions $Z\left( f_{c1}\right) =Z\left(
f_{c2}\right) $ and $Z^{\prime }\left( f_{c1}\right) =Z^{\prime
}\left( f_{c2}\right) =0$ (see Fig.s 1,2). Its endpoint $A_3$ (see Fig. 1),
is characterized, like in MCT,  by one more condition:
$f_{c1}=f_{c2}=$ $f_c$ with $Z^{\prime \prime }\left( f_c\right)
=0$, corresponding to $\lambda = 1$.

\begin{figure}[ht]
\begin{center}
\includegraphics*[scale=.43]{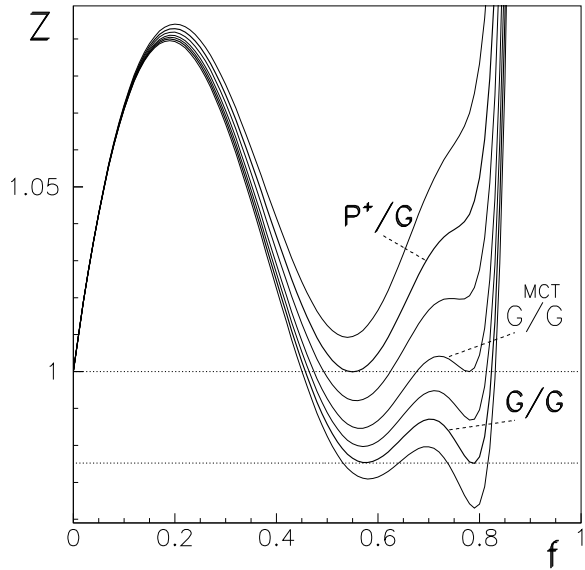}
\end{center}
\vspace{-0.9cm}
\caption{\footnotesize
The function $Z\left( f\right)$ defined in Eq.(\ref{zf}) is plotted for
several decreasing temperatures $T$ (from the top to the bottom) and fixed
density. For high $T$  the liquid solution $(P^{+})$ is given 
by the value $f=0$ corresponding to the lowest value of $Z$ namely $Z=1$.
By lowering $T$ the glass transition $T_c$ is reached when the
minimum of $Z(f)$ approaches the level $Z=1$ (bold line marked $P^+/G$). 
For $T\leq T_c$ the plateau $f_c$ is given by the value of $f$ 
where $Z(f)$ has the minimum. For low enough $T$, $Z(f)$ develops a
second minimum. The $G/G$ transition is reached when the two minima have the 
same depth (bold line marked $G/G$).
Note that in MCT approximation this would be reached instead when the 
second minimum approaches the level $Z=1$ (line marked $G/G^{MCT}$).
}
\vspace{-0.5cm}
\end{figure}


In the aging region, the correlator $C_{AG}\left( t,t^{\prime
}\right) =\Phi_{AG} \left( t,t^{\prime }\right) d$ and the
response function $G_{AG}\left( t,t^{\prime }\right)$ obey a
generalized FDT relation, $TG_{AG}\left( t,t^{\prime }\right)=
x\partial C_{AG}\left( t,t^{\prime }\right) / \partial t^{\prime
}$, where $x\leq 1$ is a constant with the physical meaning of
the ratio between the bath temperature and an effective
temperature $T_{eff}$, describing the out of equilibrium
configurations $x = T/T_{eff}$ \cite{review}.

{\em The dynamics along the G/G line.} 
On the $G/G$ line two glasses coexist corresponding to two
different values of the plateau, $f$, and thus two different values 
of $x=x(f)$ and $\lambda=\lambda(f)$  (with
$0\leq x\leq 1$ and $0\leq \lambda\leq 1$), except at the endpoint 
$A_3$ where the glasses coincide.  
As shown below, along the line the value of $\lambda$ determines 
the properties of the dynamics in the FDT regime, namely the approach 
to the plateau, while the ratio $\lambda /x$ determines the aging 
regime, i.e., the departure from the plateau. 
Four different behaviors are found along the G/G line (see Fig. 1): 
the case $\lambda /x<1$, corresponding to the points 
from $C$ to $B_3$ (resp. $B_3'$, on the other side of the line, see below); 
the new dynamical singularity points $B_3$ and $B_3'$ where 
$\lambda /x =1$; the case $\lambda /x>1$ extending from $B_3$ 
(resp. $B_3'$) to the endpoint $A_3$; and the singular point $A_3$
itself where $\lambda /x>1$, but $\lambda=1$. 

More precisely the approach to the plateau in the FDT regime,
along the entire G/G line, is given by $\Phi_{FDT} \left( \tau
\right) -f_{c}\sim \tau ^{-\beta}$, where the exponent
$\beta$ is given by $\Gamma ^2\left(1-\beta \right) /\Gamma
\left(1-2\beta \right)= \lambda $ \cite{review,CD,CL} ($\Gamma $ is
Euler gamma function). At the endpoint $A_3$ where $\lambda =1$
this is replaced by a logarithmic behavior given by $\Phi_{FDT}
\left( \tau \right) -f_c\sim 1/\ln ^{2}\tau $ \cite{gotze,MCT,gotze1}.

The departure from the plateau in the aging regime, along the G/G
line from the crossing point $C$ up to the two new points $B_3$
and $B_3'$, one for each side of the line, is given by a power
law:
\begin{equation}
\Phi_{AG} \left( t^{\prime }+\tau ,t^{\prime } \right) -f_{c}\sim
-\left( \tau /{\cal T}\left( t^{\prime }\right) \right)^{\alpha }
\end{equation}
with $\Gamma ^2\left(1+\alpha \right) /\Gamma \left(1+2\alpha
\right)= \lambda/x$ \cite{CD}. Here the $t^{\prime }$-dependence
is contained in ${\cal T}\left( t^{\prime }\right)
\sim t' $ \cite{notaeta}.

At the two new dynamical singular points, $B_3$ and
$B_3'$, where $\lambda /x=1$, this power law is replaced by a logarithmic
behavior, one for each side of the $G/G$ line.  For instance, by
approaching $B_3$ from the right (the side where the plateau value
$f_c$ is higher) one has
\begin{equation}
\Phi_{AG} \left( t^{\prime }+\tau ,t^{\prime }\right) -f_c\sim 1/\ln
^{2}
\left( \tau /{\cal T}\left( t^{\prime }\right)
\right)~~.
\end{equation}

From the point $B_3$ to the endpoint $A_3$ (analogously from
$B_3'$ to $A_3$ on the other side of the $G/G$ line),  $\lambda
/x>1$. This implies that the present one-step Replica Symmetry Breaking
solution should not hold \cite{CL} and, instead, a spin glass-like
aging dynamics should be found in such a region \cite{review}.


Interestingly the length of the $G/G$ line can be varied and, in a
version of the model with three $p$-values \cite{CCNfuture}, the
$A_3$ endpoint singularity  
can be made to coincide with the crossing point $C$
(see Fig. 1). In this particular case the $A_3$ becomes an $A_4$
singularity, as also found by MCT \cite{MCT}
(the simplest model with such a singularity is found to be
$p_1=3,p_2=4,p_3=11$ \cite{CCNfuture}). In this case the FDT
regime is logarithmic $\Phi_{FDT} \left( \tau \right) -f_c\sim
1/\ln \tau $, as in MCT \cite{gotze,MCT,gotze1},
as well as the aging regime:
\begin{equation}
\Phi_{AG} \left( t^{\prime }+\tau ,t^{\prime }\right) -f_c\sim 1/\ln
\left( \tau /{\cal T}\left( t^{\prime }\right) \right) .
\end{equation}

{\em Spinodal lines.}
We show now that the glass transition curve $P^{+}/G$ intersects the fluid
spinodal line determining the existence of a glass spinodal line.
The two spinodals are obtained by enforcing the vanishing local stability
conditions in Eq. (\ref{d-eq}):
\begin{equation}
\beta ^2\varphi ^{\prime \prime }\left( d\right) +2\beta \chi ^{\prime
\prime }\left( d\right) -\frac 1{\left( 1-d\right) ^2}-\frac 1{d^2\left(
1-f\right) ^2}=0
\end{equation}
The $P^{-}/P^{+}$-spinodal corresponds to a solution, $T(d)$, with $f=0$,
as for the $G$-spinodal solution the value of $f$ of the glassy phase
is given by the
marginal stability condition $Z^{\prime}\left( f\right) =0$. The
two lines, completing the dynamical phase diagram,
together with their range of validity are shown in Fig. 1.

In conclusion, the model we have introduced is a simple extension of
$p$-spin glass models \cite{N1,SZ,CD} 
for the glass transition. Our model exhibits a glass/glass transition
line which can be analytically investigated. Actually, it turns out to
be different from the $G/G$ line derived by equilibrium MCT and to be
characterized by new dynamical singularity points and 
off-equilibrium properties. These results can help to shed deeper light on
the properties of attractive glasses, such as colloidal suspensions.

Work supported by MIUR-PRIN 2002, FIRB 2002, CRdC-AMRA, INFM-PCI,
EU MRTN-CT-2003-504712.


\vspace{-0.5cm}
\begin{thebibliography}{40}
\vspace{-0.5cm}
\bibitem{gotze}  W. G{\"o}tze, in {\it Liquids, freezing and glass transition},
Hansen J.P., Levesque D., Zinn-Justin editors (North Holland, Amsterdam,
1991); W. G{\"o}tze and M. Sperl, Phys. Rev. E 66 (2002) 11405; 
J. Phys. Condens. Matter 15(11) (2003) S869; M. Sperl, cond-mat 0310772

\bibitem{MCT}  K. Dawson {\em et al.}, 
Phys. Rev. E 63 (2001) 11401; 
E. Zaccarelli, G. Foffi, K. Dawson, F. Sciortino and P. Tartaglia,
Phys. Rev. E 63 (2001) 31501;

\bibitem{exp1}  F. Mallamace {\em et al.}, 
Phys. Rev. Lett. 84 (2000) 5431;
W.R. Chen, S.H. Chen and F.
Mallamace, Science 300 (2003) 619

\bibitem{exp2}  T. Eckert and E. Bartsch, Phys. Rev. Lett. 89 (2002) 125701;
K.N. Pham {\em et al.}, 
Science 296 (2002) 104; 
W.C.K. Poon, K.N. Pham, S.U. Egelhaaf and P.N. Pusey, J. Phys.
Condens. Matter 16 (2003) S269

\bibitem{review}  J.P. Bouchaud, L.F. Cugliandolo, J. Kurchan and M. Mezard,
'Out of equilibrium dynamics in spin glasses and other glassy systems' in
{\it Spin glasses and random fields}, edited by A.P. Young (World
Scientific, Singapore, 1997)

\bibitem{N1}  Th. M. Nieuwenhuizen, Phys. Rev. Lett. 74 (1995) 4289; S.
Ciuchi and A. Crisanti, Europhys. Lett. 49 (2000) 754

\bibitem{SZ}  H. Sompolinsky and A. Zippelius, Phys. Rev. B 25 (1982) 6860;
T.R. Kirkpatrick and D. Thirumalai, Phys. Rev. B 36 (1987) 5388

\bibitem{CD}  L.F. Cugliandolo and J. Kurchan, Phys. Rev. Lett. 71 (1993)
173;

\bibitem{ANS} J. Arenzon, M. Nicodemi and M. Sellitto, J. Phys. I
France 6 (1996) 1143; {\em ibid.} 7 (1997) 945

\bibitem{BEG}  A. Caiazzo, A. Coniglio and M. Nicodemi, Phys. Rev. E 66
(2002) 46101; Europhys. Lett. 65(2) (2004) 256

\bibitem{CL} L.F. Cugliandolo and P. Le Doussal, Phys. Rev. E 53 (1996) 1525

\bibitem{nota1}  The full MCT equation also contains a second-order
time derivative term \cite{gotze}. Since it becomes negligible at
large times, it is often ignored in literature, as explained in
\cite{review}.

\bibitem{gotze1}  W. G{\"o}tze and M. Sperl, Phys. Rev. E 66 (2002) 11405; 
cond-mat 0403278. M. Sperl, cond-mat 0310772

\bibitem{CCNfuture} detailed calculations will be reported in a longer
paper by A. Caiazzo, A. Coniglio and M. Nicodemi, in preparation

\bibitem{KH}  H. Kinzelbach and H. Horner, J. Phys. I (France) 3 (1993)
1329; J. Phys. I (France) 3 (1993) 1901; S. Franz and M. Mezard, Physica A
209 (1994) 1; A. Crisanti, H. Horner and H.J. Sommers, Z. Phys. B 92 (1993)
257
\bibitem{notaeta} More precisely ${\cal T}\left( t^{\prime }\right)=
h\left(t^{\prime }\right) /h^{\prime }\left( t^{\prime
}\right)$,where $h\left( t\right) $ is a time reparametrization
increasing function (for the $p$-spin glass model one can assume $h\left(
t\right) \sim t$, see \cite{CD}).



\end{thebibliography}
\end{document}